\documentclass{PoS}

\title{Nucleon isovector axial charge in 2+1-flavor domain-wall QCD with physical mass}

\ShortTitle{Nucleon axial charge in  domain-wall QCD with physical mass}

\author{\speaker{Shigemi Ohta} for the LHP, RBC, and UKQCD Collaborations\\
        Institute of Particle and Nuclear Studies, KEK, Tsukuba, Ibaraki, 305-0801, Japan\\
        Department of Particle and Nuclear Studies, SOKENDAI, Hayama, Kanagawa, 240-0193, Japan\\
        RIKEN-BNL Research Center, BNL, Upton, NY, 11973, USA\\
        E-mail: \email{shigemi.ohta@kek.jp}}

\abstract{
Nucleon isovector vector, \(g_V\), and axialvector, \(g_A\), charges calculated on a 2+1-flavor dynamical domain-wall-fermions (DWF) ensemble at physical mass jointly generated by RIKEN-BNL-Columbia (RBC) and UKQCD Collaborations with lattice cut off of 1.730(4) GeV, are reported with about a percent statistical errors, along with isovector ``scalar,'' \(g_S\), and ``tensor charges,'' \(g_T\), with larger statistical errors.
Nucleon mass is estimated as 947(6) MeV.
A few standard-deviation systematics is seen in the vector charge, likely from \(O(a^2)\) discretization error through small excited-state contamination.
The axialvector charge is found with a few to several standard-deviation systematic deficit, depending on calculation methods, in comparison with the experiment.
Nucleon signal is likely lost as early as 10 lattice units or about 1.1 fm in time from the source.

\vspace{-171mm}\parbox{\textwidth}{\flushright\large\rm \hfill KEK-TH-2082, RBRC-1293}\vspace{164mm}
}

\FullConference{The 36th Annual International Symposium on Lattice Field Theory - LATTICE2018\\
		22-28 July, 2018\\
		Michigan State University, East Lansing, Michigan, USA.}

\begin{document}

\section{Introduction}

The RBC and UKQCD Collaborations have been jointly generating 2+1-flavor dynamical DWF lattice-QCD ensembles for a time \cite{Blum:2000kn,Aoki:2004ht,Allton:2008pn,Aoki:2010dy,Arthur:2012yc,Blum:2014tka}.
As is well known, DWF maintains continuum-like flavor and chiral symmetries up to \(O(a^2)\) discretization effects. 
The joint Collaborations reached the physical mass \cite{Blum:2014tka} about five years ago, and since then have been producing interesting results on pion and kaon physics and muon anomalous magnetic moment.
The joint Collaborations also have been working on nucleon structure using these 2+1-flavor dynamical DWF ensembles \cite{Yamazaki:2008py,Yamazaki:2009zq,Aoki:2010xg,Lin:2014saa,Ohta:2013qda,Ohta:2014rfa,Ohta:2015aos,Abramczyk:2016ziv,Ohta:2017gzg}.
Recently the LHP Collaboration started working with the joint Collaborations in calculating nucleon structure using the physical-mass DWF ensembles \cite{Syritsyn:2014xwa}.
Here the current status of nucleon isovector vector, \(g_V\),  and axialvector, \(g_A\), charges calculated on a physical-mass 2+1-flavor dynamical domain-wall fermions (DWF) lattice-QCD ensemble at lattice cut off of about 1.730(4) GeV  \cite{Blum:2014tka} are reported, along with isovector ``scalar charge,'' \(g_S\), and transversity, or ``tensor charge,'' \(g_T\).

Standard DWF local-current definitions are used for the quark isovector bilinears.
The required non-perturbative renormalizations for vector, axialvector, and scalar bilinears have been worked out except for tensor, in the meson-sector \cite{Blum:2014tka}.
The present result for the vector-charge renormalization provides a further opportunity to scrutinize this renormalization.
In contrast, the calculation of the axialvector charge, experimentally known as \(g_A/g_V = 1.2724(23)\) \cite{PhysRevD.98.030001}, has been more problematic:
The RBC Collaboration have been reporting  deficit in \(g_A\) \cite{Yamazaki:2008py,Yamazaki:2009zq,Aoki:2010xg,Lin:2014saa}, and values such as about 1.15(5) were reported in their most recent and lightest ensembles \cite{Ohta:2013qda,Ohta:2014rfa,Ohta:2015aos,Abramczyk:2016ziv,Ohta:2017gzg}.
Possible causes of this systematics have been extensively discussed, such as excited-state  contaminations and finite lattice volume.
The RBC Collaboration were yet to see any evidence that their vector and axialvector charges suffer any excited-state contamination.
These observations appear to have been confirmed by several other major collaborations \cite{Dragos:2016rtx,Bhattacharya:2016zcn,Liang:2016fgy,Ishikawa:2018rew,Chang:2018uxx} using different actions but with similar lattice spacings and quark masses, though extrapolations to physical mass seem to differ.
Especially important for calculations with Wilson-fermion quarks \cite{Dragos:2016rtx,Bhattacharya:2016zcn,Ishikawa:2018rew} was to remove the \(O(a)\) discretization systematic errors \cite{Liang:2016fgy}. 
The present calculation at the physical mass, however, suggests it may have captured some excited-state contamination as will be discussed in the following.
On the other hand it is well known from the days of the MIT bag model \cite{Chodos:1974pn} that the so-called ``pion cloud'' around nucleon would be important for this observable, and proper account of its geometry \cite{Adkins:1983ya}.

\section{Numerics}

We use the ``48I'' \(48^3\times 96\) 2+1-flavor dynamical M\"{o}bius DWF ensemble at physical mass with Iwasaki gauge action of \(\beta=2.13\), or of lattice cut off of  \(a^{-1} = 1.730(4)\) GeV, jointly generated by the RBC and UKQCD Collaborations\cite{Blum:2014tka}.
In total 130 configurations, separated by 20 MD trajectories in the range of trajectory number 620 to 980 and by 10 MD trajectories in the range of trajectory number from 990 to 2160, except the missing 1050, 1070, 1150, 1170, 1250, 1270, and 1470, are used.
Each configuration is deflated with 2000 low Dirac eigenvalues \cite{Clark:2017wom}.
The ``AMA'' statistics trick  \cite{Shintani:2014vja}, with \(4^4=256\) AMA sloppy samples unbiased by 4 (in time) precision ones from each configuration, is used.
The standard single-elimination jackknife is applied to the 130 samples thus obtained.
The same gauge-invariant Gaussian smearing  \cite{Alexandrou:1992ti,Berruto:2005hg} with similar parameters as in the past RBC nucleon structure calculations, are applied to nucleon source and sink, separated by \(8 \le T \le 12\) in time.

\section{Results}

\begin{figure}[t]
\begin{center}
\includegraphics[width=.49\textwidth,clip]{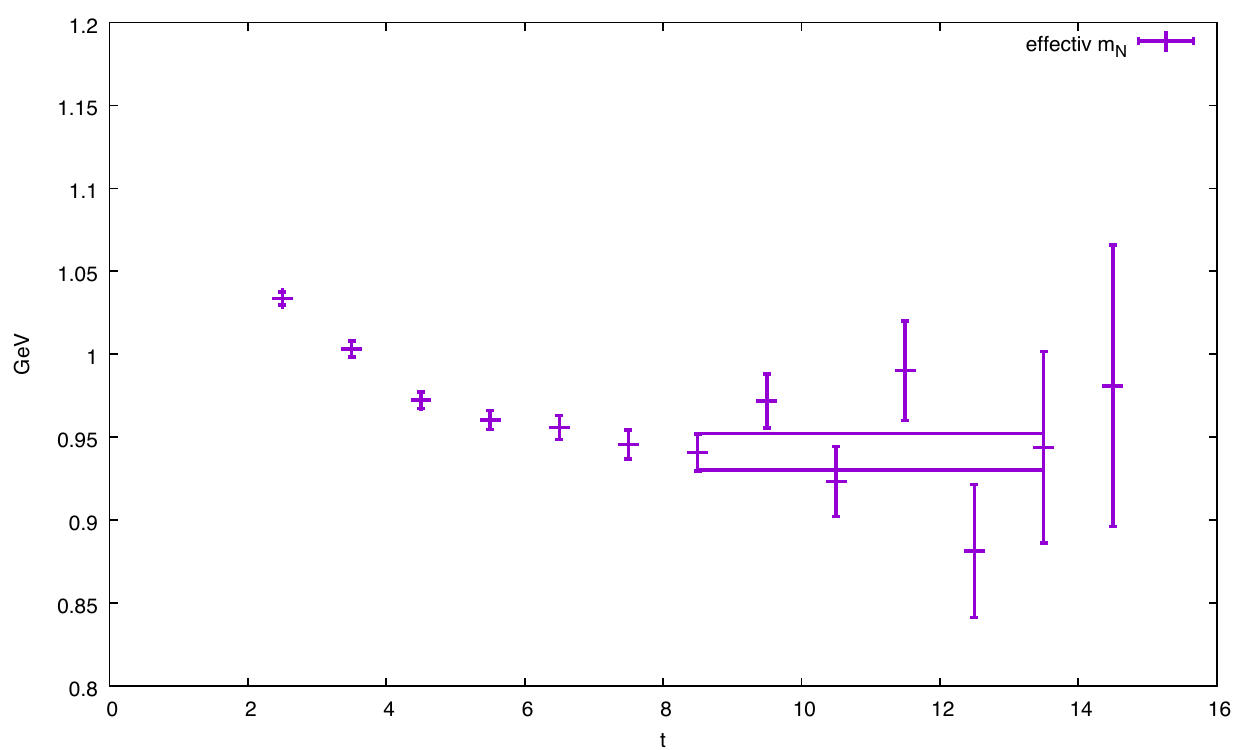}
\includegraphics[width=.49\textwidth,clip]{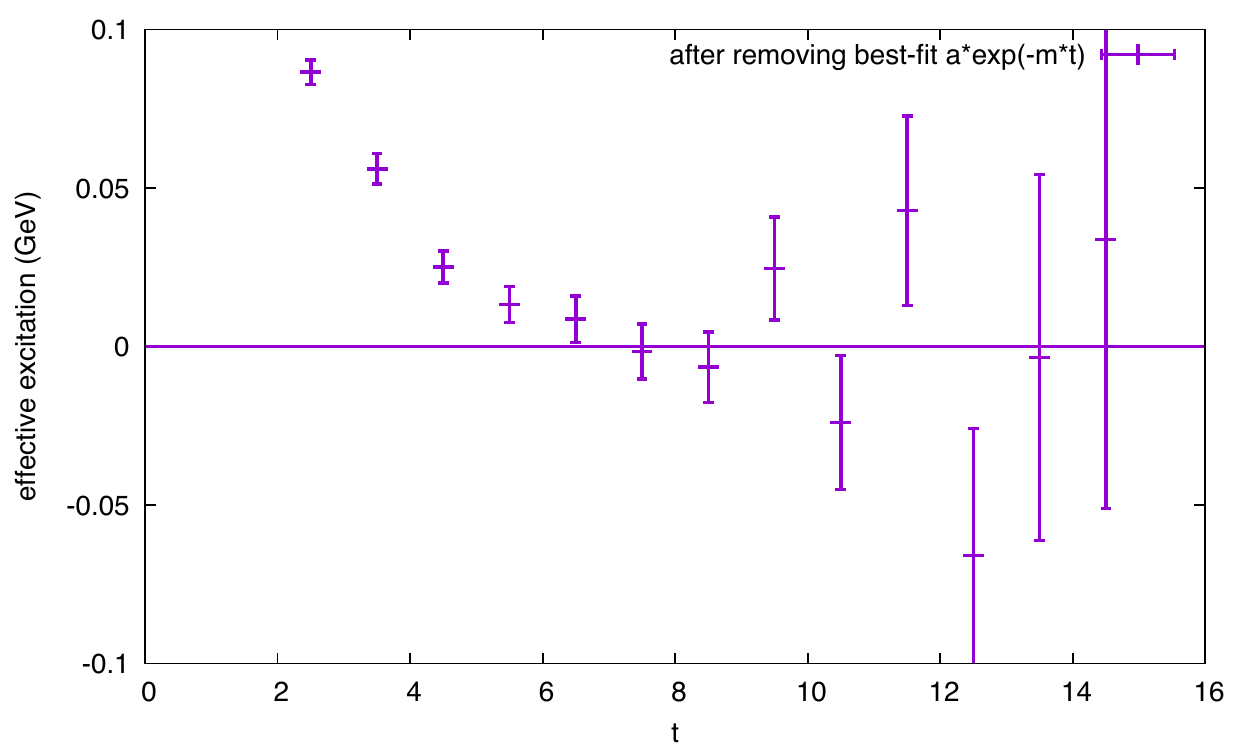}
\end{center}
\caption{\label{fig:mN}
On the left: Nucleon effective mass from the RBC+UKQCD 48I ensemble.
From a single exponential best-fit between T=8 and 13, \((6.3(2)\times10^{-9}) \times \exp(-0.547(3)t)\), we get a nucleon mass estimate of 947(6) MeV.
On the right are the signal after dividing by the best fit, to guess excited-state contamination.}
\end{figure}

The nucleon effective mass results are presented in Fig.\ \ref{fig:mN}.
From this a fit range of T=8--13 was determined to obtain a best-fit nucleon two-point correlator as \((6.3(2)\times10^{-9}) \times \exp(-0.547(3)t)\).
The nucleon mass of 0.547(3) lattice units corresponds to 947(6) MeV.
On the right pane of the figure are the excited-energy signals left after dividing by the best-fit correlator.
The excited states can be separated in the range \(t \le 7\).
We decided to use source-sink separation in time of \(8 \le T \le 12\) for the nucleon structure calculation.

\begin{figure}[t]
\begin{center}
\includegraphics[width=.49\textwidth,clip]{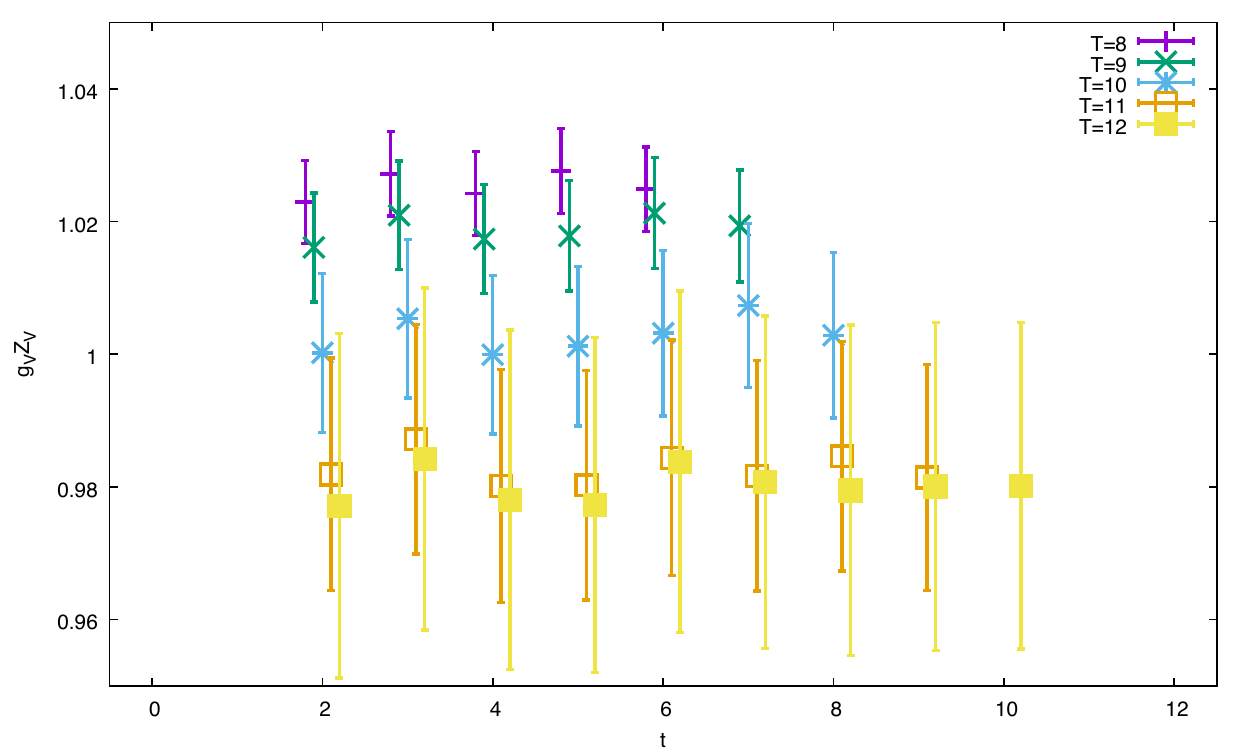}
\includegraphics[width=.49\textwidth,clip]{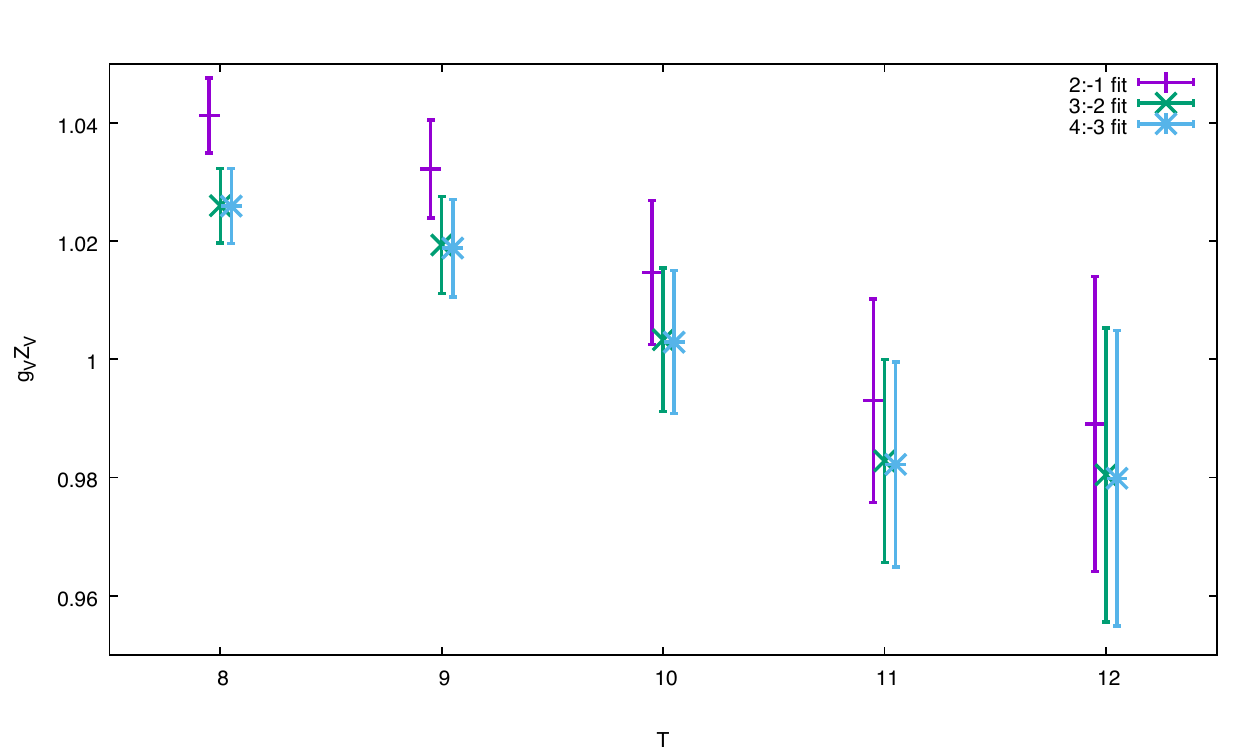}
\end{center}
\caption{\label{fig:gV}
Nucleon isovector vector charge, \(g_V\),from the RBC+UKQCD 48I ensemble, with renormalization, \(Z_V = 0.71076(25)\) obtained in the meson sector \cite{Blum:2014tka}.
The left pane presents the plateaux from source-sink separations of \(T = 8\), 9, 10, 11, and 12, respectively.
The right pane presents the fits to the plateaux with removal of first and last 2, 3, and 4 points close to the source or sink, plotted against the source-sink separation, \(T\).
Note the statistical errors of these plateau fits are about a percent or less.
}
\end{figure}

The results for the isovector vector charge, \(g_V\), obtained from the local current, are presented in Fig.\ \ref{fig:gV}, with renormalization, \(Z_V = 0.71076(25)\) obtained in the meson sector \cite{Blum:2014tka}.
On the left are plotted plateaux for five source-sink separations of \(T = 8\), 9, 10, 11, and 12, respectively.
On the right are plotted fits to these plateaux with elimination of first and last two, three, and four points close to the source or sink.
The statistical errors of these plateau fits are about a percent or less.

As can be seen the results with first and last two-point elimination seem to systematically deviate from fits with first and last three and four-point elimination, while the latter two agree well with each other.
Though not quite statisitically significant, this deviation likely signals leak through the DWF fifth dimension.
The results from the source-sink separations of \(T=8\) and 9 deviate from unity by a few percent, and also by a few standard deviations, while the results from the longer separations capture unity within their respective standard deviations.

These deviations can be attributed to excited-state contamination through discretization that is expected at \(O(a^2)\), or a couple of percent.
For this to happen there must be some excited states present in the smeared nucleon source or sink with small amplitude.
A higher-dimensional \(O(a^2)\) discretization term in the local current that is not diagonal between the ground and excited states picks up such a small-amplitude contamination of the excited state to result in what we see as the deviation from unity here.
If confirmed, this would be the first time we see such contamination in this quantity, the charge of the vector current.
However seemingly steeper slope at the separation \(T\) of 9--10 than of 8--9 prevents us from firmly concluding so, as a single excited state that is more rapidly decaying than the ground state would result in a shallower slope at larger separation.

A possible explanation for the steeper slope at larger separation is that the nucleon ground-state signal itself is beginning to disappear there, by \(T=10\): larger statistical errors here and also in the nucleon effective mass in Fig.\ \ref{fig:mN} in this time range suggest this as the likeliest cause.

Calculations at shorter source-sink separations such as \(T=7\) and 6 would be useful, so are planned in the nearest future.
The separations shorter than 6 are impractical given the amount of fifth-dimensional leak.
Until such shorter-separation calculations are performed, the data from the larger separations of \(T=10\), 11, and 12 will stay rather useless.

\begin{figure}[t]
\begin{center}
\includegraphics[width=.49\textwidth,clip]{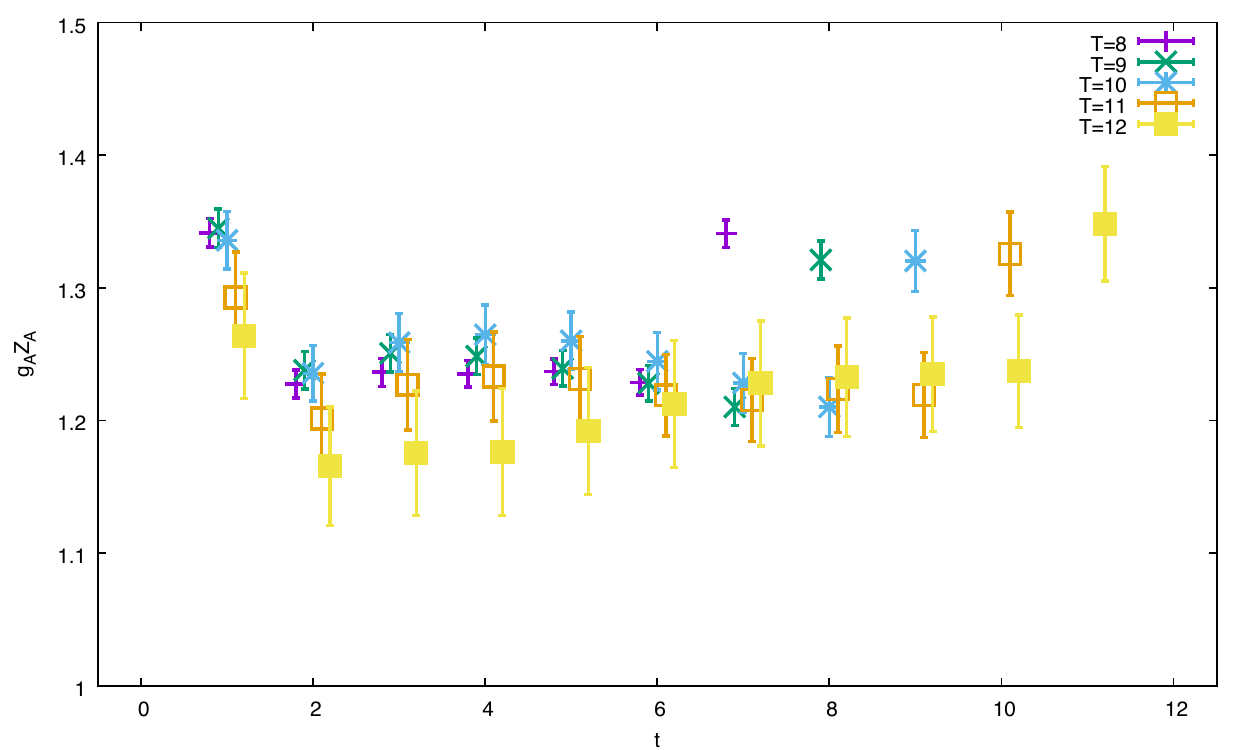}
\includegraphics[width=.49\textwidth,clip]{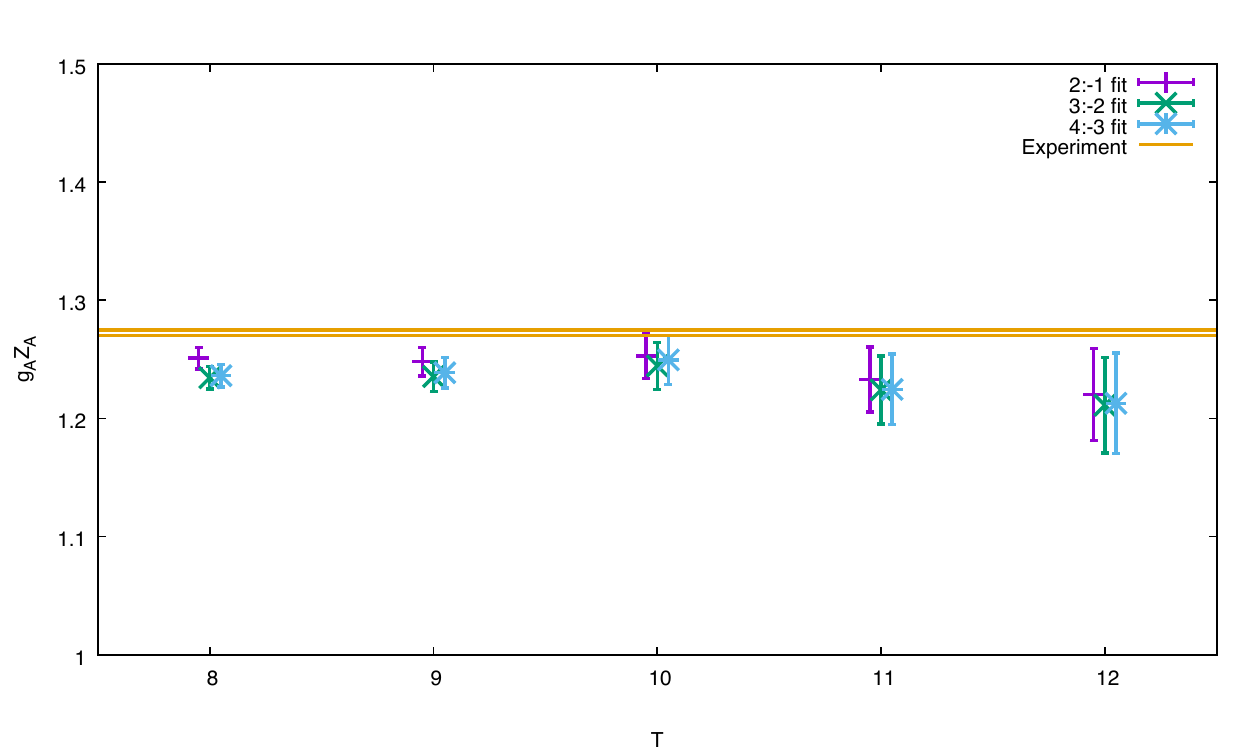}
\end{center}
\caption{\label{fig:gA}
Nucleon isovector axialvector charge, \(g_A\),from the RBC+UKQCD 48I ensemble, with renormalization, \(Z_A = 0.71191(5)\) obtained in the meson sector \cite{Blum:2014tka}.
The left pane presents the plateaux from source-sink separations of \(T = 8\), 9, 10, 11, and 12, respectively.
The right pane presents the fits to them with removal of first and last 2, 3, and 4 points close to the source or sink, plotted against the source-sink separation, \(T\).
Note the statistical errors of these plateau fits are about a percent or less.}
\end{figure}

The results for the isovector axialvector charge, \(g_A\), obtained from the local current, are presented in Fig.\ \ref{fig:gA}, with renormalization, \(Z_A = 0.71191(5)\) obtained in the meson sector \cite{Blum:2014tka}.
On the left are plotted plateaux for five source-sink separations of \(T = 8\), 9, 10, 11, and 12, respectively.
On the right are plotted fits to these plateaux with elimination of first and last two, three, and four points close to the source or sink.
Again the statistical errors of these plateau fits are about a percent or less.

Here again a slight systematic difference between plateau fits with eliminations of first and last two points from the ones with first and last three or four points is seen.
From the latter fits that agree well with each other, a slight deficit compared with the experimental value is observed, at a few standard-deviation.
In contrast to the vector charge, here with the axialvector charge there is no appreciative dependence on the source-sink separation either:
a few-standard-deviation deficit appears solid.

\begin{figure}[b]
\begin{center}
\includegraphics[width=.49\textwidth,clip]{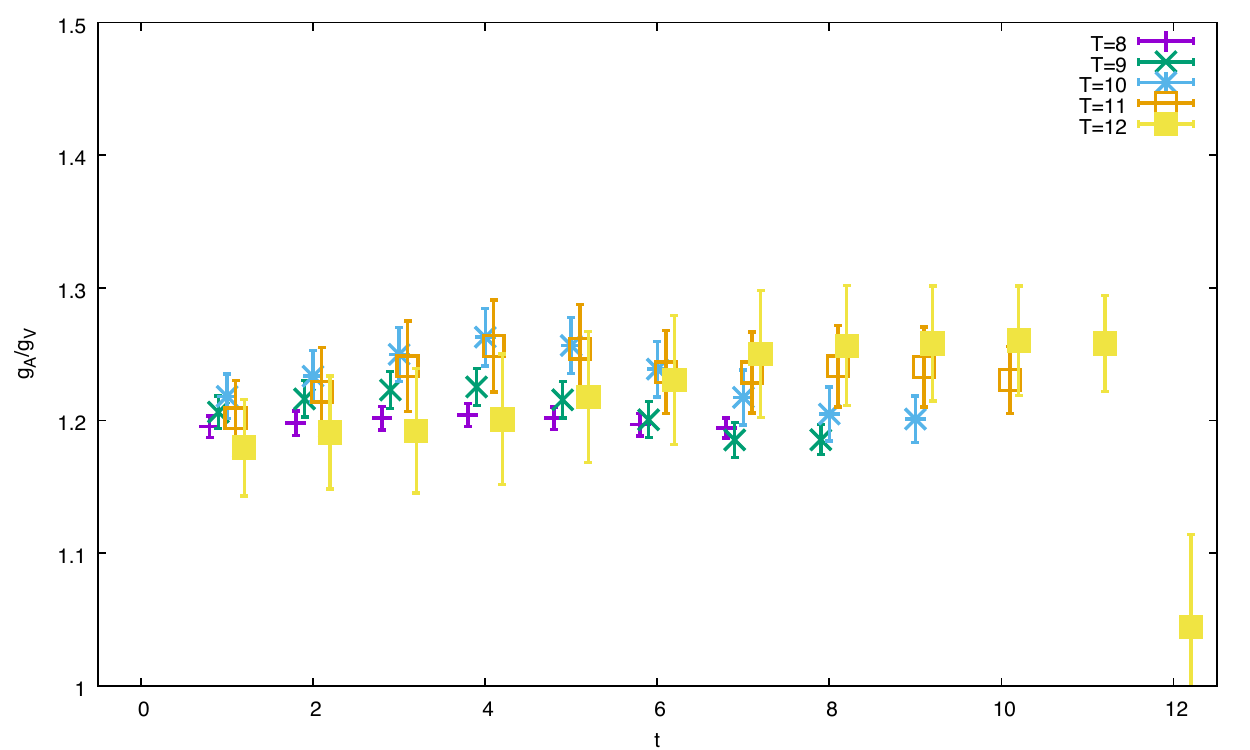}
\includegraphics[width=.49\textwidth,clip]{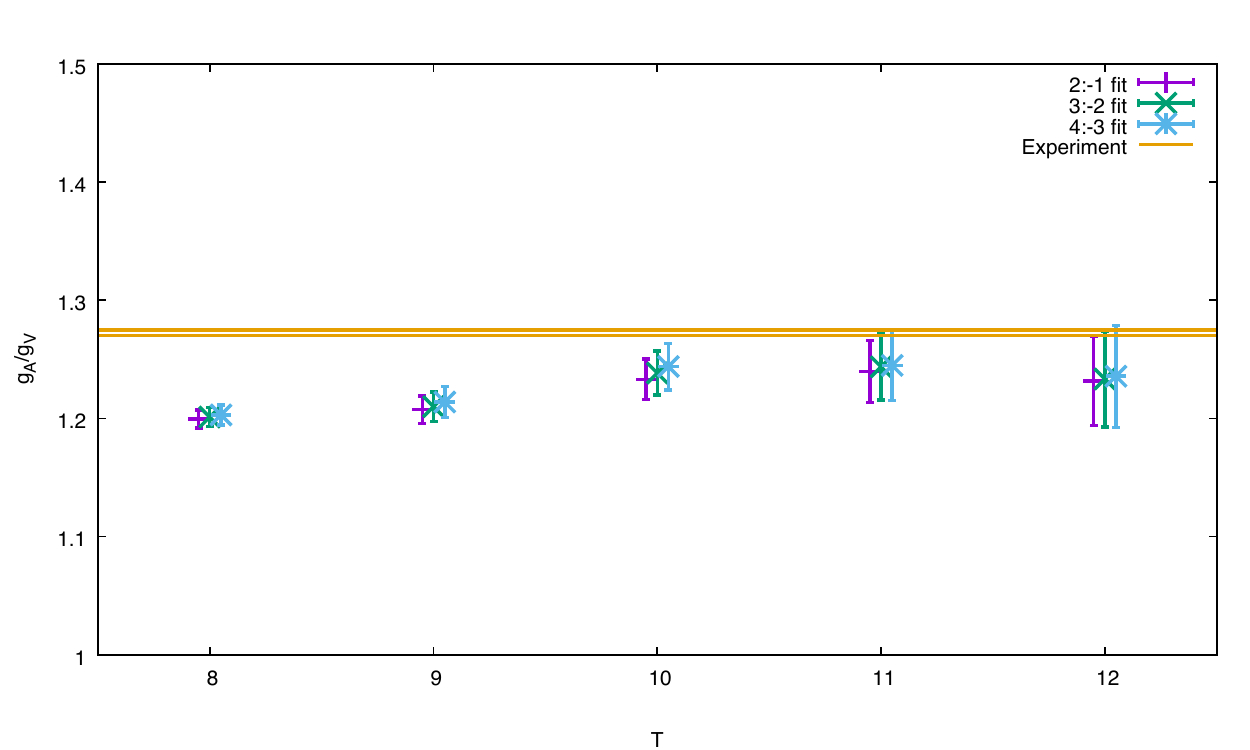}
\end{center}
\caption{\label{fig:AV}
Ratio, \(g_A/g_V\), of nucleon isovector axialvector to vector charges from the RBC+UKQCD 48I ensemble \cite{Blum:2014tka}, obtained by taking direct ratio of respective three-point correlation functions.
The left pane presents the plateaux from source-sink separations of \(T = 8\), 9, 10, 11, and 12, respectively.
The right pane presents the fits to them with removal of first and last 2, 3, and 4 points close to the source or sink, plotted against the source-sink separation, \(T\).}
\end{figure}

For the ratio, \(g_A/g_V\), of the isovector axialvector and vector charges, we can use another method of calculation by directly taking the ratios of the respective three-point correlation functions, without involving the nucleon two-point correlation function or renormalization obtained in the meson sector.
The results obtained by this method are presented in Fig.\ \ref{fig:AV}.
With this method the systematics arising from plateau selection disappears, but the steeper slope at separation \(T\) of 9--10 than of 8--9 reemerges.
And the deficit from the experimental value at shorter separation grows to several standard deviation.

It is not easy to determine the cause of this deficit in comparison with the experiment, nor the cause of difference between the two calculation methods.
However the planned additional calculations with shorter separation of \(T=7\) or 6 in the nearest future should help, especially by deciding if excited-state contamination can cause the difference of the two calculation methods.


The results for the bare isovector transversity, or ``tensor charge,'' is much noisier than that for the vector or axialvector charges in the above,
\begin{figure}[t]
\begin{center}
\includegraphics[width=.49\textwidth,clip]{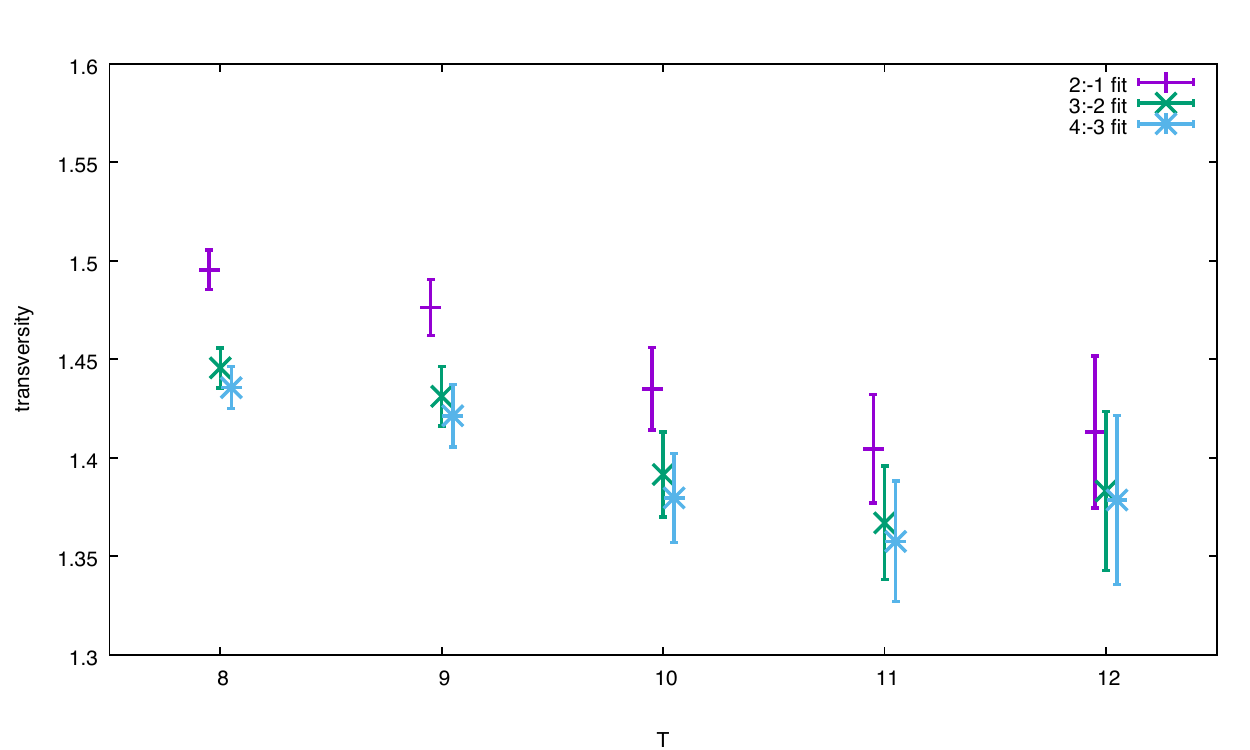}
\includegraphics[width=.49\textwidth,clip]{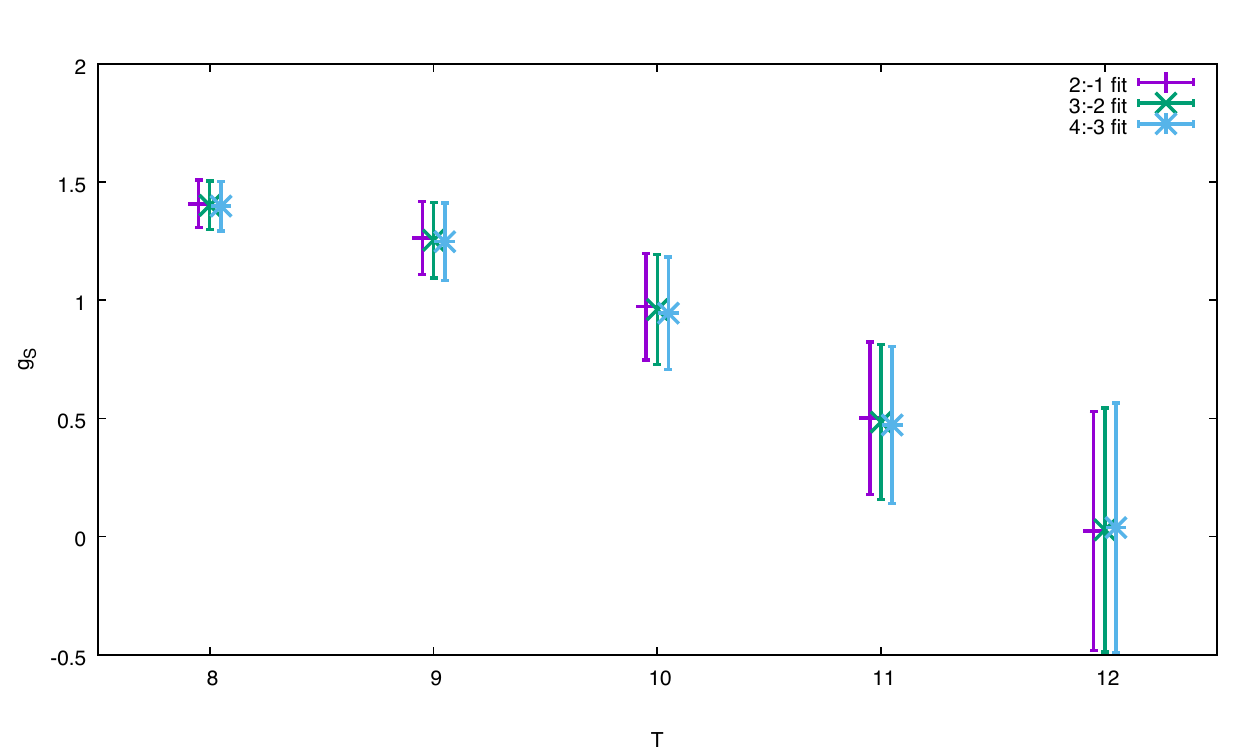}
\end{center}
\caption{\label{fig:TS}
Left: isovector transversity from  the RBC+UKQCD 48I ensemble \cite{Blum:2014tka}, unrenormalized yet.
Right: isovector ``scalar charge'' from  the RBC+UKQCD 48I ensemble \cite{Blum:2014tka}, unrenormalized though we know the renormalization.}
\end{figure}
as can be seen in the left pane of Fig.\ \ref{fig:TS}.
Here we see systematics from plateau selection.
Also a steeper slope at separation \(T\) of 9--10 than of 8--9 is seen, again suggesting the loss of nucleon signal.
The isovector ``scalar charge'' is even noisier, and perhaps because of that noise systematics from plateau selection appears absent.
A strong dependence on the separation, \(T\) and yet again steeper slope at \(T\) of 9--10 than of 8--9, is seen.

\section{Summary}

Mass and isovector vector and axialvector charges of nucleon are calculated using a physical-mass 2+1-flavor dynamical DWF lattice-QCD ensemble at a single lattice cut off of about 1.730(4) GeV:
A nucleon mass estimate of 947(6) MeV is obtained, with less than a percent of statistical error.
With similarly small statistical error of around one percent, the isovector vector charge, \(g_V\), is found to deviate from its value estimated in the meson sector, by a few percent and by a few standard deviation.
This deviation most likely arises from the expected \(O(a^2)\) discretization effect.
We will immediately add some auxiliary calculations at shorter source-sink separations to study this in detail.
In contrast our longer source-sink separations beyond \(T \ge 10\) lattice units may be useless, as the nucleon signal, with our current statistics, likely are dying there.
The isovector axialvector charge, \(g_A\), is calculated again with similarly small statistical error of about a percent.
It is found in deficit from the experimental value by a few to several standard deviations depending on the calculation methods.
This will again be followed up immediately by auxiliary calculations with shorter source-sink separations.
Isovector transversity, or ``tensor charge,'' \(g_T\), and ``scalar charge,'' \(g_S\), are also calculated with behaviors consistent with the above observation that the nucleon signal may be dying by the source-sink separation of about ten lattice units.

The author thanks the members of LHP, RBC, and UKQCD Collaborations, and in particular Sergey Syritsyn.
The ``48I'' ensemble was generated using the IBM Blue Gene/Q (BG/Q) ``Mira'' machines at the Argonne Leadership Class Facility (ALCF) provided under the Incite Program of the US DOE, on the ``DiRAC'' BG/Q system funded by the UK STFC in the Advanced Computing Facility at the University of Edinburgh, and on the BG/Q  machines at the Brookhaven National Laboratory.
The nucleon calculations were done using ALCF Mira.
The author is partially supported by Japan Society for the Promotion of Sciences, Kakenhi grant 15K05064.

\bibliographystyle{epj}
\bibliography{nucleon}

\end{document}